\documentclass[twocolumn,amsmath,amssymb,a4paper,prb,superscriptaddress,floatfix]{revtex4-1}
\usepackage[dvipdfmx]{graphicx}
\usepackage{natbib}
\usepackage{multirow}
\usepackage{amsmath}
\usepackage{bm}
\usepackage{mathrsfs}
\usepackage{url}
\usepackage{color}
\usepackage{ulem}
\begin{document}

\title{Phonon structure of titanium under shear deformation
along $\{10\bar{1}2\}$ twinning mode}

\author{Atsushi Togo}
\email{togo.atsushi@gmail.com}
\affiliation{
Research and Services Division of
Materials Data and Integrated System, National Institute for Materials Science,
Tsukuba, Ibaraki 305-0047, Japan}
\affiliation{Center for Elements Strategy Initiative for Structural
Materials, Kyoto University, Sakyo, Kyoto 606-8501, Japan}

\author{Yuta Inoue}
\affiliation{Department of Materials Science and
Engineering, Kyoto University, Sakyo, Kyoto 606-8501, Japan}

\author{Isao Tanaka}
\affiliation{Center for Elements Strategy Initiative for Structural
Materials, Kyoto University, Sakyo, Kyoto 606-8501, Japan}
\affiliation{Department of Materials Science and
Engineering, Kyoto University, Sakyo, Kyoto 606-8501, Japan}
\affiliation{Nanostructures Research Laboratory, Japan Fine Ceramics
Center, Atsuta, Nagoya 456-8587, Japan}

\begin{abstract}
 We investigated phonon behavior of hexagonal close packed titanium
 under homogeneous shear deformation corresponding to the
 $\{10\bar{1}2\}$ twinning mode using first-principles calculation and
 phonon calculation. By this deformation, we found that a phonon mode
 located at a point on Brillouin zone boundary is drastically soften
 increasing the shear and finally it triggers a spontaneous structural
 transition by breaking the crystal symmetry toward twin from parent.
\end{abstract}

\maketitle

\section{Introduction}
The crystallography of deformation twinning has been studied by Bilby
{\it et al.}\cite{Bilby-1965} in 1965 and Bevis {\it et
al.}\cite{Bevis-1968} in 1968. Croker and Bevis also reported that for
titanium specifically in 1970.\cite{Croker-1970} These studies were well
summarized in the review paper by Christian and Mahajan in
1995.\cite{Christian-1995} These studies are based on the concept that
the deformation twinning is characterized by collective displacement of
atoms.\cite{Cayron-2018} Therefore, phonon should play a central role
in the deformation twinning. Recently, increase of computational power
has enabled to perform first principles phonon calculation routinely. We
applied this phonon calculation to the deformation twinning to discuss
the structural transition from parent to twin in a quantitative manner
as described in the following sections. In this report, we take the
$\{10\bar{1}2\}$ twinning mode of the hexagonal close-packed (HCP)
titanium as a simple example to apply our general approach using the
phonon calculation.

There are recent reports by using first principles calculations on
microscopic mechanism of deformation twinning involving twin
boundary.\cite{Pei-2017,Jiang-2019} However we take the traditional
crystallographic approach to construct the calculation model with
homogeneous shear without including the twin boundary. The
first-principles phonon calculation is applied to this sheared-parent
structure model to seek a characteristic phonon mode that exhibits
structural instability under a certain shear. We show that the
spontaneous structural transition induced by this phonon instability
triggers transformation from the sheared-parent to twin.  The
final collective displacement of atoms predicted by this calculation
excellently agrees with those reported in
Refs.~\onlinecite{Bilby-1965,Bevis-1968, Croker-1970,Christian-1995} as
the shuffling mechanism.

Throughout this report, we employ symmetry constraints to distinguish
the parent, sheared-parent, shuffling, sheared-twin, and twin
structures. Here the shuffling structure is defined as the instantaneous
structure that appears during rearrangement of atomic positions by the
transformation from the sheared-parent to sheared-twin structure. More
details are provided in the following sections. The aim of use of
crystal symmetry is to analyze the mechanism of the deformation twinning
in a comprehensible way by limiting the number of degrees of atomistic
freedom. Thus an important contribution from this report may be detailed
symmetry information provided along the transition pathway from the
parent to twin structure.

This report is organized as follows. In
Sec.~\ref{sec:calculation-model}, a calculation model of the crystal
structure is defined using a transformation matrix to represent the
homogeneous shear corresponding to the $\{10\bar{1}2\}$ twinning
mode. In Sec.~\ref{sec:method-of-calculation}, computational details of
the first principles calculation and phonon calculation are given along
with the convergence test against $\mathbf{k}$-point density and
smearing width of electronic structure calculation. In
Sec.~\ref{sec:results-and-discussion}, the transformation from the
parent to twin structure is discussed from a viewpoint of collective
displacement of atoms. Finally, shuffling mechanism of the
deformation twinning is associated with symmetry breaking of the
sheared-parent structure induced by the unstable phonon mode.

\section{Calculation model}
\label{sec:calculation-model}

\subsection{Unit cell representations}
\begin{figure}[ht]
  \begin{center}
    \includegraphics[width=0.85\linewidth]{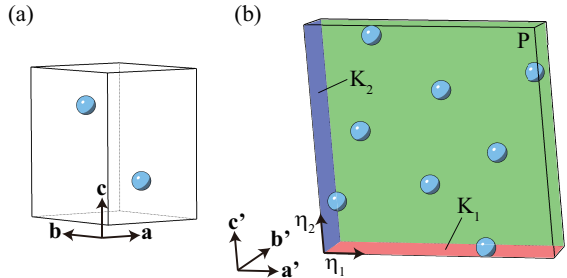}
   \caption{(color online) Unit cell structures of HCP titanium, (a) the
   conventional unit cell and (b) an extended unit cell for easy
   representation of the $\{10\bar{1}2\}$ twinning mode. $\mathrm{K}_1$
   (red) and $\mathrm{K}_2$ (blue) planes and $\eta_1$ and $\eta_2$
   directions characterize the twinning mode. The plane of shear is
   depicted by $\mathrm{P}$ (green). \label{fig:HCP-twin-structures} }
  \end{center}
\end{figure}

The HCP crystal structure is shown in Fig.~\ref{fig:HCP-twin-structures}
(a). The set of the basis vectors is given as
$(\mathbf{a},\mathbf{b},\mathbf{c})$. To represent the $\{10\bar{1}2\}$
twinning mode characterized by $\eta_1$ and $\eta_2$ directions,
$\mathrm{K}_1$ and $\mathrm{K}_2$ planes, and the plane of shear
$\mathrm{P}$, it is convenient to retake the HCP unit cell by its
extended unit cell as depicted in Fig.~\ref{fig:HCP-twin-structures}
(b). The set of the basis vectors of the extended unit cell,
$(\mathbf{a}',\mathbf{b}',\mathbf{c}')$, is chosen to satisfy the
conditions $\eta_1 \parallel \mathbf{a}'$, $\eta_2 \parallel
\mathbf{c}'$, and $\mathbf{b}' \perp \mathrm{P}$. The change-of-basis is
easily represented using an integer matrix $\mathrm{Q}$:
\begin{align}
 \label{eq:change-of-basis}
 (\mathbf{a}',\mathbf{b}',\mathbf{c}') &=
 (\mathbf{a},\mathbf{b},\mathbf{c}) \mathrm{Q}
\end{align}
with
\begin{align}
 \mathrm{Q} &=
  \begin{pmatrix}
   2 & 0 & \bar{2} \\
   1 & 1 & \bar{1} \\
   1 & 0 & 1
  \end{pmatrix},
\end{align}
where the bars on the matrix elements denote the negative numbers. The
conditions of $\mathbf{b}' \perp \mathbf{a}'$ and $\mathbf{b}' \perp
\mathbf{c}'$ are easily confirmed finding $(2\mathbf{a} + \mathbf{b})
\cdot \mathbf{b} = 0$.

\subsection{Lattice under homogeneous shear}
\begin{figure}[ht]
  \begin{center}
    \includegraphics[width=0.50\linewidth]{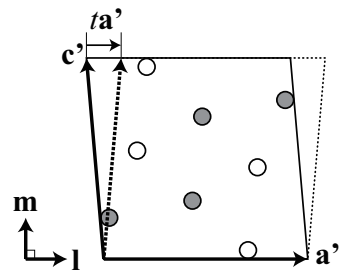}
   \caption{Homogeneous shear corresponding to the $\{10\bar{1}2\}$
   twinning mode. The extended unit cell given by
   Eq.~(\ref{eq:change-of-basis}) is projected on the plane of shear
   $\mathrm{P}$. The atoms depicted by the same symbols (open or filled)
   are located on the same plane parallel to the plane of shear. These
   two planes for the atoms with the open and filled symbols are shifted
   by $0.5\mathbf{b}'$ each other. The lattice drawn by the dotted lines
   is the sheared lattice. Applying the shear, $\mathbf{c}'$ is changed
   to $\mathbf{c}' + t\mathbf{a}'$ with $t$ as a representative
   parameter. \label{fig:twin-structure-2D} }
  \end{center}
\end{figure}

Following the usual crystallographic definition of deformation
twinning,\cite{Christian-1995} we impose homogeneous shear
corresponding to the $\{10\bar{1}2\}$ twinning mode to the extended unit
cell. In Fig.~\ref{fig:twin-structure-2D}, the extended unit cell is
illustrated being projected on the plane of shear $\mathrm{P}$. The
basis vectors $\mathbf{a}'$ and $\mathbf{c}'$ are parallel to the
$\mathrm{P}$ plane, and the basis vector $\mathbf{b}'$ is perpendicular
to the $\mathrm{P}$ plane.

The homogeneous shear is written by an affine transformation:
\begin{align}
 \label{eq:affine-transformation}
\mathbf{v} = \mathrm{S} \mathbf{u},
\end{align}
where $\mathrm{S}$ is the matrix representation of the shear, and
$\mathbf{u}$ and $\mathbf{v}$ are the vectors.
By this homogeneous shear $\mathrm{S}$, the basis
vectors $(\mathbf{a}', \mathbf{b}', \mathbf{c}')$ are transformed
to $(\mathbf{a}'_\mathrm{s}, \mathbf{b}'_\mathrm{s},
\mathbf{c}'_\mathrm{s})$ as follows:
\begin{align}
 \label{eq:basis-transformation-by-tensor}
 \mathbf{a}'_\mathrm{s} = \mathrm{S} \mathbf{a}',\;
 \mathbf{b}'_\mathrm{s} = \mathrm{S} \mathbf{b}',\;
 \mathbf{c}'_\mathrm{s} = \mathrm{S} \mathbf{c}'.
\end{align}

Following the convention by Christian and Mahajan in
Ref.~\onlinecite{Christian-1995}, to represent the $\{10\bar{1}2\}$
twinning mode, we take a Cartesian axis $\mathbf{l}$ the unit vector
parallel to $\eta_1$ and $\mathbf{m}$ the unit normal to the
$\mathrm{K}_1$ plane as shown in
Fig.~\ref{fig:twin-structure-2D}. Choosing the orientation by
$\mathbf{l}=(1,0,0)^\mathrm{T}$ and $\mathbf{m}=(0,1,0)^\mathrm{T}$, the
basis vectors $(\mathbf{a}', \mathbf{b}', \mathbf{c}')$ are given in the
Cartesian coordinates as
\begin{align}
 \label{eq:cartesian-basis}
 \mathbf{a}' =
 \begin{pmatrix}
  a'_x \\  0 \\ 0
 \end{pmatrix},
 \mathbf{b}' =
 \begin{pmatrix}
  0 \\ 0 \\ b'_z
 \end{pmatrix}
 \mathbf{c}' =
 \begin{pmatrix}
  c'_x \\ c'_y \\ 0
 \end{pmatrix}.
\end{align}
By this choice of the coordinates, the homogeneous shear $\mathrm{S}$ is
simply represented by only one free variable
$s$~\cite{Christian-1995},
\begin{align}
 \label{eq:matrix-S}
 \mathrm{S} =
 \begin{pmatrix}
   1 & s & 0 \\
   0 & 1 & 0 \\
   0 & 0 & 1
 \end{pmatrix}.
\end{align}
The basis vectors $\mathbf{a}'$ and $\mathbf{b}'$ are unchanged applying
this $\mathrm{S}$, whereas $\mathbf{c}'$ is transformed to
$\mathbf{c}'_\mathrm{s} = (c'_x + sc'_y, c'_y, 0)^\mathrm{T}$.

It is more convenient for us to represent the homogeneous shear as a
linear combination of the basis vectors since the choice of the
orientation in the Cartesian coordinates becomes arbitrary. This
is written by a transformation matrix $\mathrm{T}$:
\begin{align}
 \label{eq:basis-transformation}
(\mathbf{a}'_\mathrm{s},\mathbf{b}'_\mathrm{s},\mathbf{c}'_\mathrm{s}) =
(\mathbf{a}',\mathbf{b}',\mathbf{c}') \mathrm{T}.
\end{align}
To have $\mathbf{c}'_\mathrm{s}=\mathbf{c}' + t\mathbf{a}'$ as shown in
Fig.~\ref{fig:twin-structure-2D}, we take
\begin{align}
 \mathrm{T} =
 \begin{pmatrix}
   1 & 0 & t \\
   0 & 1 & 0 \\
   0 & 0 & 1
 \end{pmatrix}.
\end{align}
In the Cartesian coordinates (\ref{eq:cartesian-basis}), the relationship
between $s$ and $t$ is given as
\begin{equation}
 \label{eq:twinning-shear-relationship}
 sc'_y=ta'_x.
\end{equation}
This is found by combining
Eqs.~(\ref{eq:basis-transformation-by-tensor}) and
(\ref{eq:basis-transformation}) as $\mathrm{SL} = \mathrm{LT}$, where
$\mathrm{L}$ is the $3\times 3$ matrix whose columns are
$\mathbf{a}'$, $\mathbf{b}'$, and $\mathbf{c}'$ in
the Cartesian coordinates~(\ref{eq:cartesian-basis}).

From Eqs.~(\ref{eq:change-of-basis}) and
(\ref{eq:basis-transformation}), we obtain the basis vectors under the
homogeneous shear,
$(\mathbf{a}_\mathrm{s},\mathbf{b}_\mathrm{s},\mathbf{c}_\mathrm{s})$, by
\begin{align}
 \label{eq:deformed-HCP-unitcell}
 (\mathbf{a}_\mathrm{s},\mathbf{b}_\mathrm{s},\mathbf{c}_\mathrm{s}) =
 (\mathbf{a},\mathbf{b},\mathbf{c}) \mathrm{Q} \mathrm{T} \mathrm{Q}^{-1}.
\end{align}
Here Eq.~(\ref{eq:deformed-HCP-unitcell}) is independent on the
orientation of Cartesian coordinates. The combined transformation
matrix $\mathrm{QTQ^{-1}}$ is used to construct the sheared HCP unit
cells for the first-principles calculation and phonon calculation.

\subsection{Sheared-parent structure and shuffling}
\label{sec:sheared-parent-shuffling}

At a certain twinning shear $s=s_\mathrm{t}$ in the matrix
(\ref{eq:matrix-S}), the sheared lattice gets to have the same lattice
as the parent lattice with a different orientation, and thus the same
point group symmetry as the parent lattice. The twinning shear of the
$\{10\bar{1}2\}$ twinning mode is known as $s_\mathrm{t}=\frac{3 -
\gamma^2}{\sqrt{3}\gamma}$ with
$\gamma=|\mathbf{c}|/|\mathbf{a}|$.\cite{Bilby-1965, Bevis-1968,
Croker-1970} Similarly, that in $t$ is easily found from
Fig.~\ref{fig:twin-structure-2D} as $t_\mathrm{t}=-2\mathbf{c}'\cdot
\mathbf{a}'/|\mathbf{a}'|^2 = \frac{2(3 - \gamma^2)}{3 + \gamma^2}$.
Sicne the ratio $s/s_\mathrm{t}=t/t_\mathrm{t}$ is a good measure of the
homogeneous shear in this study, we employ $s/s_\mathrm{t}$ as the
representative parameter in the following part of this report.

Even at $s/s_\mathrm{t}=1$, the space group symmetry of the
sheared-parent structure is unnecessarily the same as that of the parent
structure. For the sheared-parent structure to have the same space group
symmetry as the parent, i.e., to form the twin, usually a certain
rearrangement of the atomic positions is required. This rearrangement
may be called ``shuffling'', and for the use of this terminology clearly
in our calculations, we define it in a little more detail as follows.

The space group type of the HCP crystal structure is $P6_3/mmc$
(No. 194) and its site symmetry type (Wyckoff position) of all atoms is
$6\bar{m}2$ ($c$). Obviously, the homogeneous shear for the
$\{10\bar{1}2\}$ twinning mode breaks these symmetries. Keeping
crystallographic coordinates of atoms unchanged under the lattice shear,
these symmetries are reduced to $C2/m$ (No. 12) and $m$ ($i$),
respectively, except at $s/s_\mathrm{t}=1$, $Cmcm$ (No. 63) and $m2m$
($c$), respectively.

For simplicity, we assume that sheared structure is either
sheared-parent structure or sheared-twin structure in the interval of
$\{0 \leq s/s_\mathrm{t} \leq 1\}$, and the sheared-parent and
sheared-twin structures may transform each other by rearranging their
atomic positions. In this rearrangement, we expect that the atoms are
displaced collectively and their displacement distances are much shorter
than the distance between the nearest neighboring atoms.

We fix the crystal symmetry of the
sheared-parent structure as $C2/m$. Unless breaking this symmetry, we
allow to relax the positions of the atoms, by which the atoms can be
displaced within their planes parallel to the plane of shear
$\mathrm{P}$. The atomic displacement $\mathbf{u}(s/s_\mathrm{t})$ by
this relaxation is written as
\begin{align}
 \label{eq:position-relaxation}
 \mathbf{u}(s/s_\mathrm{t}) =& \mathbf{r}(s/s_\mathrm{t}) - \mathrm{S}
 \mathbf{r}(0)
 \nonumber \\
 =& (\mathbf{a}',\mathbf{b}',\mathbf{c}')
 (\boldsymbol{x}(s/s_\mathrm{t}) -
 \mathrm{T} \boldsymbol{x}(0)),
\end{align}
where $\mathbf{r}(s/s_\mathrm{t})$ is the position of the atom after the
relaxation at $s/s_\mathrm{t}$ and $\boldsymbol{x}(s/s_\mathrm{t})$ is
the crystallographic coordinates corresponding to
$\mathbf{r}(s/s_\mathrm{t})$. The second line of
Eq.~(\ref{eq:position-relaxation}) is the convenient expression for us
to construct the crystal structure models. This relaxation is not
considered as a part of the shuffling in this study. We require that the
shuffling arises in association with temporal symmetry breaking by the
rearrangement of the atomic positions.

\section{Method of calculation}
\label{sec:method-of-calculation}

\subsection{Computational details}
\label{computational-details}

The phonon calculations were performed by the finite displacement method
with the supercell approach, where the atoms were displaced. The
supercells were created by $4 \times 4\times 3$ multiplication of the
sheared HCP unit cells obtained by $\mathrm{QTQ^{-1}}$ in
Eq.~(\ref{eq:deformed-HCP-unitcell}). For each phonon calculation,
atomic displacements of 0.02 \AA~were introduced to the perfect
supercells along $\mathbf{a}_\mathrm{s}$ and $\mathbf{c}_\mathrm{s}$ in
plus and minus directions individually. For the phonon calculation, we
employed the phonopy code.\cite{phonopy}

For the first-principles calculations, we employed the plane-wave basis
projector augmented wave method~\cite{PAW-Blochl-1994} within the
framework of density functional theory (DFT) as implemented in the VASP
code.\cite{VASP-Kresse-1995,VASP-Kresse-1996,VASP-Kresse-1999} The
generalized gradient approximation of Perdew, Burke, and Ernzerhof
revised for solids~\cite{PBEsol} was used as the exchange
correlation potential. A plane-wave energy cutoff of 300 eV was
employed. The radial cutoff of the PAW dataset of Ti was 1.323 \AA. The
3{\it{p}} electrons for Ti were treated as valence and the remaining
electrons were kept frozen. The smearing method was used with the
$k$-point sampling on a uniform mesh for the Brillouin zone integration
of the electronic structure. The $8 \times 8 \times 6$ and $2 \times 2
\times 2$ $k$-point sampling meshes with half grid shifts along $c^*$
directions were used for the sheared HCP unit cells and their $4 \times
4\times 3$ supercells, respectively, in conjunction with the smearing
width $\sigma=0.4$ eV in the Methfessel-Paxton
scheme.\cite{Methfessel-1989} These parameters were chosen after
convergence check of the lattice parameter and phonon band structure as
presented in the next section.

To perform systematic calculations presented below, we employed the
AiiDA environment~\cite{AiiDA} with the AiiDA-VASP~\cite{AiiDA-VASP} and
AiiDA-phonopy~\cite{AiiDA-phonopy} plugins.

\subsection{Choices of calculation parameters}

To choose the $k$-point sampling mesh density and smearing width
$\sigma$, we examined convergence of the lattice parameter of the HCP
conventional unit cell and the phonon band structure with respect to
these values. In general, use of denser $k$-point sampling mesh provides
better calculation accuracy at a constant smearing width
$\sigma$. However, to save the computational demand, we expect to employ
a sparser $k$-point sampling mesh. For this purpose, we may choose a
larger $\sigma$ value although sacrificing fine detail of electronic
structure. We have to find a good compromise between the
$k$-point sampling mesh and smearing width $\sigma$ against the required
accuracy of our computational research.

\begin{figure}[ht]
  \begin{center}
    \includegraphics[width=0.90\linewidth]{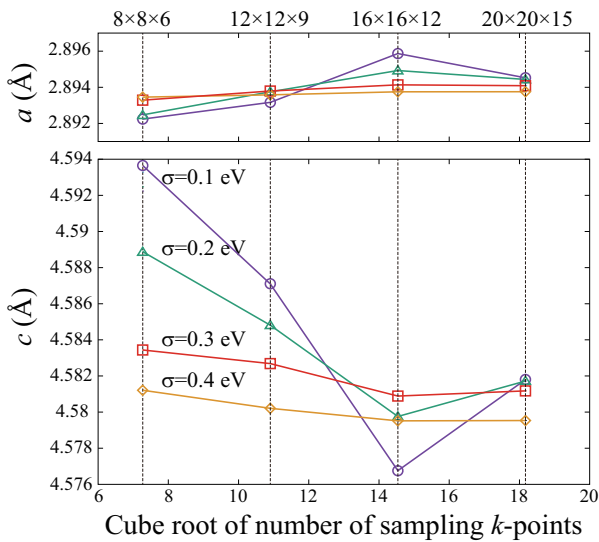}
   \caption{(color online) Calculated lattice parameters $a$ and $c$ of
   the HCP conventional unit cell with respect to $k$-point sampling
   mesh density and smearing width $\sigma$. The circle, triangle,
   square, and diamond symbols denote those calculated with
   $\sigma=0.1$, 0.2, 0.3, and 0.4 eV, respectively. The lines
   connecting the points are guide to the
   eye. \label{fig:lattice-params-conv} }
  \end{center}
\end{figure}

The calculated lattice parameters $a$ and $c$ with respect to the
$k$-point sampling mesh of $8\times 8\times 6$, $12\times 12\times 9$,
$16\times 16\times 12$, and $20\times 20\times 15$ for the HCP unit cell
and $\sigma=0.1$, $0.2$, $0.3$, and $0.4$ eV are shown in
Fig.~\ref{fig:lattice-params-conv}. Clearly, the lattice parameter $c$
is more sensitive to both of the $k$-point sampling mesh density and
$\sigma$ than $a$. At $\sigma=0.2$, $0.3$, and $0.4$ eV, the lattice
parameters are getting to converge increasing the $k$-point sampling
mesh density. Use of $\sigma=0.1$ eV requires an even denser $k$-point
sampling mesh, for which the phonon calculation with the supercell
approach can be computationally too demanding for us. Therefore, we
decided not to consider the use of $\sigma=0.1$ eV. At $\sigma=0.4$ eV,
the lattice parameters are roughly constant at the different $k$-point
sampling meshes, and $a$ and $c$ are enough close to the more accurate
results by $\sigma=0.2$ eV and $20 \times 20\times 15$ mesh. With the
parameter pair of $\sigma=0.4$ eV and $8 \times 8\times 6$ mesh, which
requires the smallest computational demand in this convergence test, we
obtained $a=2.893$ and $c=4.581$~\AA. These values underestimate the
experimental values of $a=2.951$ and $c=4.684$~\AA~\cite{Wood-1962} by a
few percent. This discrepancy is considered reasonably small for the
purpose of this study.

\begin{figure}[ht]
  \begin{center}
    \includegraphics[width=0.90\linewidth]{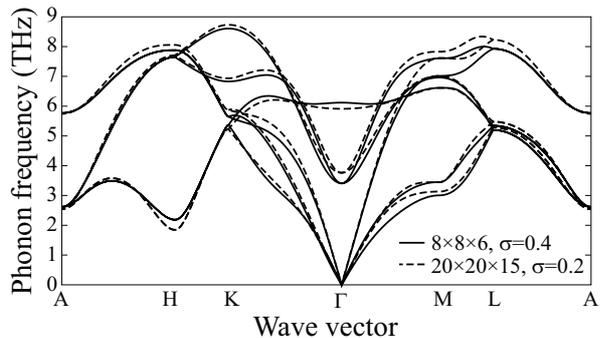}
   \caption{Phonon band structures calculated with two different
   parameter pairs for Brillouin zone integration of the electronic
   structure calculation, $8\times 8\times
   6$ $k$-point sampling mesh ($2\times 2\times 2$ mesh for supercell)
   and 0.4 eV smearing width (solid curve) and $20\times 20\times
   15$ $k$-point sampling mesh ($5\times 5\times 5$ mesh for supercell)
   and 0.2 eV smearing width (dashed curve). The coordinates of the wave
   vector labels are given in
   Ref.~\onlinecite{Aroyo-2014}. \label{fig:phonon-conv} }
  \end{center}
\end{figure}

The phonon band structure calculated with the $8\times 8\times 6$
$k$-point sampling mesh ($2\times 2\times 2$ for supercell) and
$\sigma=0.4$ eV is compared with that with the $20\times 20\times 15$
$k$-point sampling mesh ($5\times 5\times 5$ for supercell) and
$\sigma=0.2$ eV as shown in Fig.~\ref{fig:phonon-conv}. These phonon
band structures reasonably agree with the
experiment in Ref.~\onlinecite{Stassis-Ti-phonon-1979} and calculation
in Ref.~\onlinecite{Won-Seok-Ti-phonon-2015}. Although we can see
noticeable difference in phonon frequency between them, we consider this
difference is small enough to discuss characteristic behavior of the
phonon mode that we are interested in.  Therefore we chose the $k$-point
sampling mesh of $8\times 8\times 6$ along with the smearing width of
$\sigma=0.4$ eV.

\section{Results and discussion}
\label{sec:results-and-discussion}

\subsection{Optimization of sheared-parent structures}
\label{sec:sheard-parent-structures}

\begin{figure}[ht]
  \begin{center}
    \includegraphics[width=0.90\linewidth]{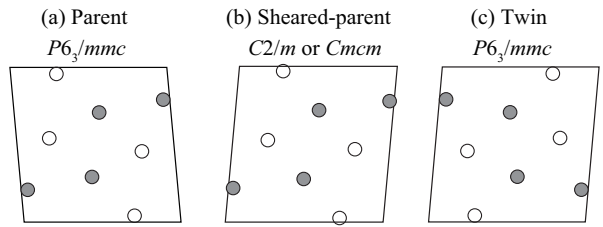}
   \caption{Crystal structures of (a) parent, (b) sheared-parent, and
   (c) twin. The space group type of the parent and twin structures is
   $P6_3/mmc$. That of the sheared-parent structure is $C2/m$ at the
   general $s/s_\mathrm{t}$ and $Cmcm$ at $s/s_\mathrm{t} = 1$. See the
   caption of Fig.~\ref{fig:twin-structure-2D} for the different symbols
   of atoms.  \label{fig:structure-evolution} }
  \end{center}
\end{figure}

The crystal structure models of the parent, sheared-parent, and twin
structures are shown in Fig.~\ref{fig:structure-evolution}. The parent
and twin structures have the space group type of $P6_3/mmc$ whereas the
twin is oriented in a different direction from the parent after the
shuffling.

\begin{figure}[ht]
  \begin{center}
    \includegraphics[width=0.90\linewidth]{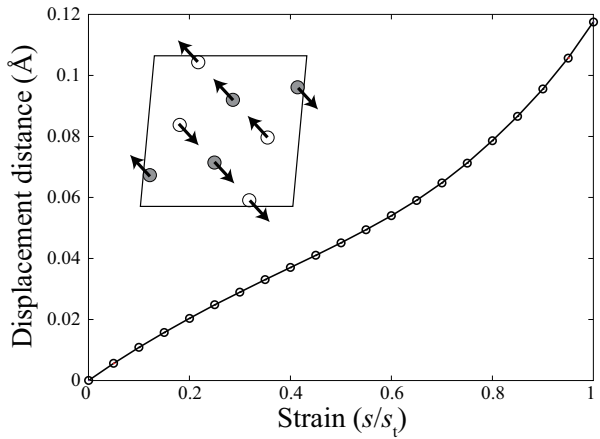}
   \caption{\label{fig:relaxation-distance} Displacement distances
   $|\mathbf{u}(s/s_\mathrm{t})|$ of Eq.~(\ref{eq:position-relaxation})
   at the homogeneous shears corresponding to the $\{10\bar{1}2\}$
   twinning mode. The lines connecting the open circles are guide to the
   eye. The sheared-parent structure at $s/s_\mathrm{t}=1$ before the
   optimization is shown in the inset.  The arrows depict approximate
   directions of the displacements. The structure after the optimization
   is shown in Fig.~\ref{fig:structure-evolution} (b). See the caption
   of Fig.~\ref{fig:twin-structure-2D} for the different symbols of
   atoms.}
  \end{center}
\end{figure}

To create the sheared-parent structure models, we sampled 21 homogeneous
shears in the interval of $\{0 \le s/s_\mathrm{t} \le 1\}$ uniformly,
and the first-principles calculations were performed to optimize their
internal atomic positions under the symmetry constraint. The
displacement distances $|\mathbf{u}(s/s_\mathrm{t})|$ (see
Eq.~(\ref{eq:position-relaxation})) obtained as
the results of the structure optimizations are shown in
Fig.~\ref{fig:relaxation-distance}. Since $|\mathbf{u}(s/s_\mathrm{t})|$
of the all atoms in the unit cell at each $s/s_\mathrm{t}$ are the same
due to the crystal symmetry, only one value at each $s/s_\mathrm{t}$ is
shown. The approximate directions of the displacements are shown in the
inset. The atoms having the same displacement directions in this figure
are symmetrically equivalent by the lattice translation, i.e., only two
directions can exist and they are directed in the opposite directions as
proposed by the crystal symmetry. The displacement distances
$|\mathbf{u}(s/s_\mathrm{t})|$ increase as increasing
$s/s_\mathrm{t}$. Even at $s/s_\mathrm{t}=1$, the displacement distance
is only about 4\% of the nearest neighbor distance. Therefore, for the
schematic analysis, performing this optimization is considered
unimportant, although it is necessary for accurate phonon calculation.

\subsection{Energy increase by homogeneous shear}

\label{sec:energy-increase}

\begin{figure}[ht]
  \begin{center}
    \includegraphics[width=0.90\linewidth]{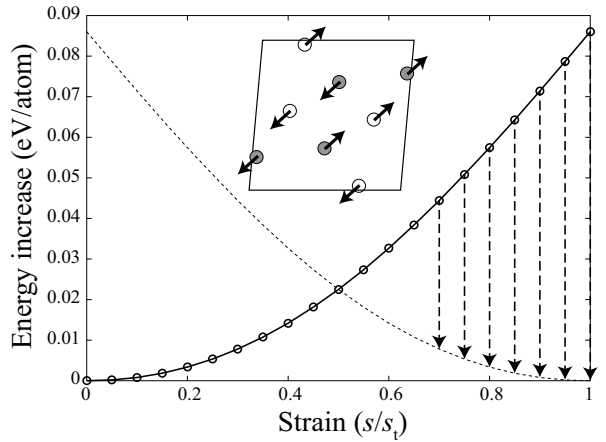}
   \caption{Increase of electronic total energy of the sheared-parent
   structure with respect to the homogeneous shear corresponding to the
   $\{10\bar{1}2\}$ twinning mode. The lines connecting the open circles
   are guide to the eye. The dotted curve is the mirror image of the
   solid curve, which represents the energy increase of the
   sheared-twin structure. The arrows with the vertical dashed lines show
   the energies that are released by the transformation from the
   sheared-parent structures to the sheared-twin structures at general
   $s/s_\mathrm{t}$ and to the twin structure at $s/s_\mathrm{t}=1$ by
   the shuffling. The inset shows the sheared-parent structure at
   $s/s_\mathrm{t}=1$. The arrows show the atomic displacement
   directions indicated by the eigenvector of the imaginary phonon mode
   at $\mathbf{q}=(1/2, 0, 0)$. See the caption of
   Fig.~\ref{fig:twin-structure-2D} for the different symbols of
   atoms.\label{fig:energy-strain} }
  \end{center}
\end{figure}

In Fig.~\ref{fig:energy-strain}, we show electronic total energy
increase of the sheared-parent structure with respect to the homogeneous
shear $s/s_\mathrm{t}$ under the symmetry constraint. Increasing the
shear, the energy increases harmonically. From the definition of the
deformation twinning, we know that the sheared-twin structures have to
give the same energy curve with respect to $1 - s/s_\mathrm{t}$. This is
drawn by the dotted curve as the mirror image of that of the parent.
The crossing of these energy curves at $s/s_\mathrm{t} = 0.5$ implies
that their energy surfaces are disjointed in the atomic configuration
space under the symmetry constraints of $C2/m$.

\subsection{Spontaneous symmetry breaking due to shear}

At $s/s_\mathrm{t} > 0.5$, breaking the symmetry constraint may allow
the sheared-parent structure to transform to the sheared-twin structure
spontaneously. This is guessed due to the fact that the $\{10\bar{1}2\}$
twinning mode exists. Therefore, we investigated this systematically as
follows.

\begin{figure}
  \begin{center}
    \includegraphics[width=1.00\linewidth]{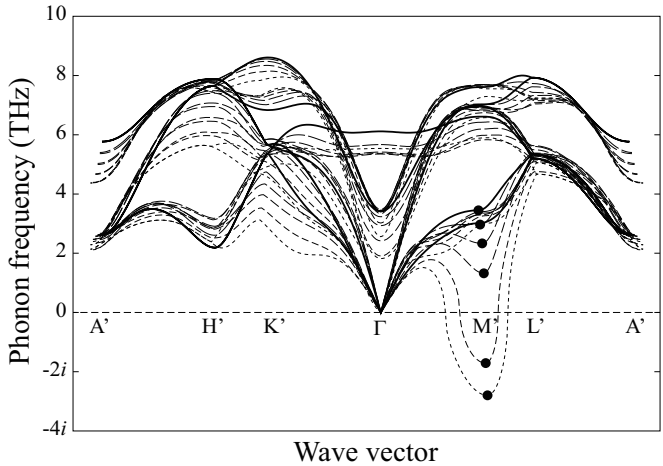}
   \caption{Phonon band structures at the homogeneous shears
   corresponding to the $\{10\bar{1}2\}$ twinning mode. Those at $s/s_t
   =$ 0, 0.2, 0.4, 0.6, 0.8, and 1 are drawn by the different line styles
   from solid to shorter dashed lines. Phonon frequency below zero means
   imaginary frequency. A similar set of labels of points in Brillouin
   zones to that in Fig.~\ref{fig:phonon-conv} is used for easy
   comparison, although the homogeneous shear breaks the crystal
   symmetry. The explicit coordinates of these points are $\mathrm{A}'
   (0, 0, 1/2)$, $\mathrm{H}' (1/3, 1/3, 1/2)$, $\mathrm{K}' (1/3, 1/3,
   0)$, $\Gamma (0, 0, 0)$, $\mathrm{M}' (1/2, 0, 0)$, and $\mathrm{L}'
   (1/2, 0, 1/2)$. Since the shapes of the Brillouin zones are different
   for the different shears, the positions of the reciprocal points in
   the Cartesian coordinates measured from the $\Gamma$ points slightly
   disagree. The filled circles at the $\mathrm{M}'$ points indicate the
   soft phonon modes that have the eigenvector shown in
   Fig.~\ref{fig:structure-evolution} (b).  \label{fig:bands-at-strains}
   }
  \end{center}
\end{figure}

We applied a series of the phonon calculations to the sheared-parent
structures at the 21 homogeneous shears that are the same shears as
those chosen in Secs.~\ref{sec:sheard-parent-structures} and
\ref{sec:energy-increase}. The phonon band structures calculated at the
selected shears are shown in Fig.~\ref{fig:bands-at-strains}. One phonon
mode at the wave vector $\mathbf{q}=(1/2, 0, 0)$ exhibits imaginary
frequency at the larger shears. This indicates that spontaneous
structural transformation should occur by breaking symmetry. Note that
the coordinates of $\mathbf{q}$ are represented in the basis
vectors given by Eq.~(\ref{eq:deformed-HCP-unitcell}).

\begin{figure}
  \begin{center}
    \includegraphics[width=0.90\linewidth]{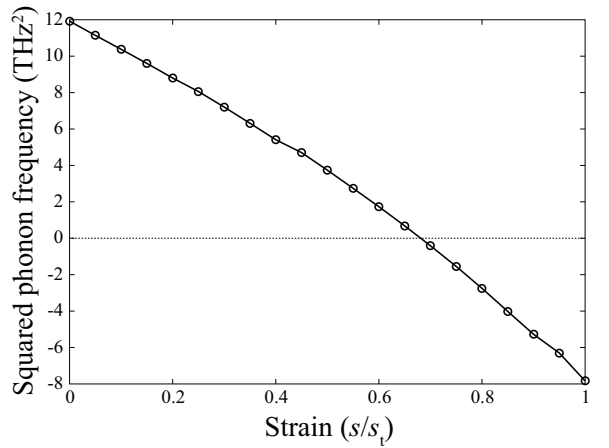}
   \caption{Squared frequencies of the characteristic phonon mode at
   $\mathbf{q}=(1/2, 0, 0)$ at the homogeneous shears
   corresponding to the $\{10\bar{1}2\}$ twinning mode. The lines
   connecting the open circles are guide to the
   eye. \label{fig:frequency2-strain} }
  \end{center}
\end{figure}

This behavior is similar to the structural phase transition induced by
changing an intensive parameter such as pressure or
temperature.\cite{IntroductionToLatticeDynamics} Assuming the
second-order-like structural phase transition, the squared frequency of
the characteristic phonon mode is considered to be related to the order
parameter.  Hence we plotted it as a function of the homogeneous shear
as shown in Fig.~\ref{fig:frequency2-strain}, and the roughly linear
trend was found as expected. From this plot, the critical shear was
estimated as $s/s_\mathrm{t} \sim 0.68$.

\subsection{Shuffling}

The polarization vector (eigenvector) of the characteristic phonon mode
at $\mathbf{q}=(1/2, 0, 0)$ contains necessary information to properly
break the crystal symmetry by introducing collective displacement of
atoms that is illustrated in the inset of
Fig.~\ref{fig:energy-strain}. In this figure, the arrows directed from
the atoms in the sheared-parent structure depict directions of the
atomic displacements with the same relative amplitude. This symmetry
breaking doubles the primitive cell and the space group type is reduced
to $P2_1/c$ (No. 15).

Small finite displacements along the polarization vector were introduced
to the sheared-parent structure models so that the first-principles
calculation code, VASP, can detect the broken symmetry correctly. Then,
we performed the first-principles calculation to optimize these
structure models at the homogeneous shears with their basis vectors
fixed. After the structure optimizations, we obtained the sheared-twin
structures at $s/s_\mathrm{t} = 0.7, \ldots, 0.95$ and the twin
structure at $s/s_\mathrm{t} = 1$, i.e., the structures did not fall
into local minima. This structural transformation is the shuffling that
we consider.

\begin{figure}
 \begin{center}
  \includegraphics[width=0.70\linewidth]{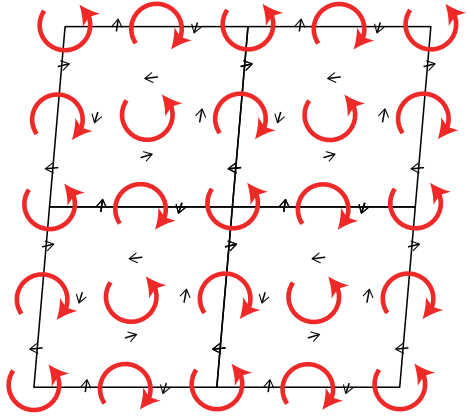} \caption{(color
    online) The small (black) arrows depict atomic displacements induced
    by the structural optimization from the sheared-parent structure
    (Fig.~\ref{fig:structure-evolution} (b)) to the twin structure
    (Fig.~\ref{fig:structure-evolution} (c)). The circular (red) arrows
    show how the displacements are arranged as structural
    units. \label{fig:shuffling} }
 \end{center}
\end{figure}

The atomic displacements by this transformation at $s/s_\mathrm{t} = 1$
is shown by small (black) arrows in Fig.~\ref{fig:shuffling}. The
displacement distance is roughly constant by $\sim$0.5 \AA~($\sim$17\%
of the interatomic distance) at all $s/s_\mathrm{t}$ larger than the
critical shear. The minimum structural unit of the collective
displacement is composed of four atoms. We can see that two types of the
parallelogram units are arranged to fill the crystal structure by
alternately changing their rotation directions depicted by the (red)
circular arrows. The spaces surrounded by these rotation units behave
like breathing. In this manner, it is considered that the internal
structural distortion required by the shuffling is minimized. This
illustration is almost the same as that obtained from the
crystallographic discussion of the shuffling for the $\{10\bar{1}2\}$
twinning mode by Crocker and Bevis in
Ref.~\onlinecite{Croker-1970}. What was found in this study is that this
shuffling necessarily occurs at the shear larger than the critical
shear.

Energy difference between the sheared-parent and sheared-twin structures
is released by the shuffling as depicted by the arrows with the vertical
dashed lines in Fig.~\ref{fig:energy-strain}. In between $s/s_\mathrm{t}
= 0.5$ and the critical shear $s/s_\mathrm{t} \sim 0.68$, the potential
energy barrier should prevent the shuffling from being initiated. In
real materials, imperfection of crystal may lower the potential energy
barrier locally, and the shuffling initiated from this point will
propagate over the whole crystal body at the velocity of sound by the
help of the released energy to result in the macroscopic
transformation. It would be difficult to observe how the twin forms from
the parent microscopically, since the sheared parent relaxes toward the
twin instantaneously. What we can measure by experiments as an evidence
of this shuffling mechanism like a structural phase transition is the
systematic frequency change of the characteristic phonon mode at
$s/s_\mathrm{t} < 0.5$ as the precursor effect, for which inelastic
neutron or X-ray scattering techniques can be employed.

\section{Summary}
We investigated the microscopic mechanism of the $\{10\bar{1}2\}$
twinning mode of the HCP titanium using the first principles calculation
and phonon calculation. As the calculation model, the sheared-parent
structures were defined by change of basis and transformation matrices
at the specific homogeneous shears corresponding to the $\{10\bar{1}2\}$
twinning mode. At the homogeneous shears of $s/s_\mathrm{t} > 0.68$, the
phonon mode instability under the symmetry constraint was exhibited at a
Brillouin zone boundary of $\mathbf{q}=(0.5, 0, 0)$. Breaking the
crystal symmetry following the polarization vector (eigenvector) of this
characteristic phonon mode, the sheared-twin structure was obtained
after optimizing internal positions of atoms. This indicates the
mechanism of the formation of the deformation twinning of this
system. These calculations provide the structural transition pathway
from the parent to twin, which may be widely accepted as the
``shuffling''. The collective displacement of atoms obtained from the
phonon calculation agrees with the shuffling mechanism reported by
Crocker and Bevis~\cite{Croker-1970} from the crystallographic argument.
We applied the same approach to the other twinning modes of HCP metals
and found that the shuffling of the $\{10\bar{1}2\}$ twinning mode is
the simplest case. For the others, it may be hard for the simple
crystallographic approach to find their reasonable collective
displacements of atoms since those atoms are rearranged in more
complicated ways three dimensionally than that of the $\{10\bar{1}2\}$
twinning mode. These results of the other twinning modes will be
published elsewhere. In this report, we provided a systematic
calculational approach to explore microscopic mechanism of deformation
twinning, and the technical details were written in this report. The
computational resource and tools necessary to perform this calculation
are usual. Therefore any researcher who is interested in the deformation
twinning can achieve the similar calculations immediately and find
specific details of their twinning modes. This is probably the largest
contribution of this report to the field of deformation twinning.

\section*{ACKNOWLEDGMENTS}
This work was supported by MEXT Japan through ESISM (Elements Strategy
Initiative for Structural Materials) of Kyoto University.

\bibliography{HCP-twinning}

\begin{thebibliography}{22}%
\makeatletter
\providecommand \@ifxundefined [1]{%
 \@ifx{#1\undefined}
}%
\providecommand \@ifnum [1]{%
 \ifnum #1\expandafter \@firstoftwo
 \else \expandafter \@secondoftwo
 \fi
}%
\providecommand \@ifx [1]{%
 \ifx #1\expandafter \@firstoftwo
 \else \expandafter \@secondoftwo
 \fi
}%
\providecommand \natexlab [1]{#1}%
\providecommand \enquote  [1]{``#1''}%
\providecommand \bibnamefont  [1]{#1}%
\providecommand \bibfnamefont [1]{#1}%
\providecommand \citenamefont [1]{#1}%
\providecommand \href@noop [0]{\@secondoftwo}%
\providecommand \href [0]{\begingroup \@sanitize@url \@href}%
\providecommand \@href[1]{\@@startlink{#1}\@@href}%
\providecommand \@@href[1]{\endgroup#1\@@endlink}%
\providecommand \@sanitize@url [0]{\catcode `\\12\catcode `\$12\catcode
  `\&12\catcode `\#12\catcode `\^12\catcode `\_12\catcode `\%12\relax}%
\providecommand \@@startlink[1]{}%
\providecommand \@@endlink[0]{}%
\providecommand \url  [0]{\begingroup\@sanitize@url \@url }%
\providecommand \@url [1]{\endgroup\@href {#1}{\urlprefix }}%
\providecommand \urlprefix  [0]{URL }%
\providecommand \Eprint [0]{\href }%
\providecommand \doibase [0]{http://dx.doi.org/}%
\providecommand \selectlanguage [0]{\@gobble}%
\providecommand \bibinfo  [0]{\@secondoftwo}%
\providecommand \bibfield  [0]{\@secondoftwo}%
\providecommand \translation [1]{[#1]}%
\providecommand \BibitemOpen [0]{}%
\providecommand \bibitemStop [0]{}%
\providecommand \bibitemNoStop [0]{.\EOS\space}%
\providecommand \EOS [0]{\spacefactor3000\relax}%
\providecommand \BibitemShut  [1]{\csname bibitem#1\endcsname}%
\let\auto@bib@innerbib\@empty
\bibitem [{\citenamefont {Bilby}\ \emph {et~al.}(1965)\citenamefont {Bilby},
  \citenamefont {Crocker},\ and\ \citenamefont {Cottrell}}]{Bilby-1965}%
  \BibitemOpen
  \bibfield  {author} {\bibinfo {author} {\bibfnamefont {B.~A.}\ \bibnamefont
  {Bilby}}, \bibinfo {author} {\bibfnamefont {A.~G.}\ \bibnamefont {Crocker}},
  \ and\ \bibinfo {author} {\bibfnamefont {A.~H.}\ \bibnamefont {Cottrell}},\
  }\href@noop {} {\bibfield  {journal} {\bibinfo  {journal} {Proc. Roy. Soc.
  A}\ }\textbf {\bibinfo {volume} {288}},\ \bibinfo {pages} {240} (\bibinfo
  {year} {1965})}\BibitemShut {NoStop}%
\bibitem [{\citenamefont {Bevis}\ \emph {et~al.}(1968)\citenamefont {Bevis},
  \citenamefont {Crocker},\ and\ \citenamefont {Rosenhead}}]{Bevis-1968}%
  \BibitemOpen
  \bibfield  {author} {\bibinfo {author} {\bibfnamefont {M.}~\bibnamefont
  {Bevis}}, \bibinfo {author} {\bibfnamefont {A.~G.}\ \bibnamefont {Crocker}},
  \ and\ \bibinfo {author} {\bibfnamefont {L.}~\bibnamefont {Rosenhead}},\
  }\href@noop {} {\bibfield  {journal} {\bibinfo  {journal} {Proc. Roy. Soc.
  A}\ }\textbf {\bibinfo {volume} {304}},\ \bibinfo {pages} {123} (\bibinfo
  {year} {1968})}\BibitemShut {NoStop}%
\bibitem [{\citenamefont {Crocker}\ and\ \citenamefont
  {Bevis}(1970)}]{Croker-1970}%
  \BibitemOpen
  \bibfield  {author} {\bibinfo {author} {\bibfnamefont {A.~G.}\ \bibnamefont
  {Crocker}}\ and\ \bibinfo {author} {\bibfnamefont {M.}~\bibnamefont
  {Bevis}},\ }in\ \href@noop {} {\emph {\bibinfo {booktitle} {The Science,
  Technology and Application of Titanium}}},\ \bibinfo {editor} {edited by\
  \bibinfo {editor} {\bibfnamefont {R.~I.}\ \bibnamefont {Jaffee}}\ and\
  \bibinfo {editor} {\bibfnamefont {N.~E.}\ \bibnamefont {Promisel}}}\
  (\bibinfo  {publisher} {Pergamon},\ \bibinfo {year} {1970})\ pp.\ \bibinfo
  {pages} {453--458}\BibitemShut {NoStop}%
\bibitem [{\citenamefont {Christian}\ and\ \citenamefont
  {Mahajan}(1995)}]{Christian-1995}%
  \BibitemOpen
  \bibfield  {author} {\bibinfo {author} {\bibfnamefont {J.~W.}\ \bibnamefont
  {Christian}}\ and\ \bibinfo {author} {\bibfnamefont {S.}~\bibnamefont
  {Mahajan}},\ }\href@noop {} {\bibfield  {journal} {\bibinfo  {journal}
  {Progress in Materials Science}\ }\textbf {\bibinfo {volume} {39}},\ \bibinfo
  {pages} {1} (\bibinfo {year} {1995})}\BibitemShut {NoStop}%
\bibitem [{\citenamefont {Cayron}(2018)}]{Cayron-2018}%
  \BibitemOpen
  \bibfield  {author} {\bibinfo {author} {\bibfnamefont {C.}~\bibnamefont
  {Cayron}},\ }\href@noop {} {\bibfield  {journal} {\bibinfo  {journal}
  {Crystals}\ }\textbf {\bibinfo {volume} {8}} (\bibinfo {year}
  {2018})}\BibitemShut {NoStop}%
\bibitem [{\citenamefont {Pei}\ \emph {et~al.}(2017)\citenamefont {Pei},
  \citenamefont {Zhang}, \citenamefont {Hickel}, \citenamefont {Fri{\'a}k},
  \citenamefont {Sandl{\"o}bes}, \citenamefont {Dutta},\ and\ \citenamefont
  {Neugebauer}}]{Pei-2017}%
  \BibitemOpen
  \bibfield  {author} {\bibinfo {author} {\bibfnamefont {Z.}~\bibnamefont
  {Pei}}, \bibinfo {author} {\bibfnamefont {X.}~\bibnamefont {Zhang}}, \bibinfo
  {author} {\bibfnamefont {T.}~\bibnamefont {Hickel}}, \bibinfo {author}
  {\bibfnamefont {M.}~\bibnamefont {Fri{\'a}k}}, \bibinfo {author}
  {\bibfnamefont {S.}~\bibnamefont {Sandl{\"o}bes}}, \bibinfo {author}
  {\bibfnamefont {B.}~\bibnamefont {Dutta}}, \ and\ \bibinfo {author}
  {\bibfnamefont {J.}~\bibnamefont {Neugebauer}},\ }\href@noop {} {\bibfield
  {journal} {\bibinfo  {journal} {Npj Comput. Mater.}\ }\textbf {\bibinfo
  {volume} {3}},\ \bibinfo {pages} {6} (\bibinfo {year} {2017})}\BibitemShut
  {NoStop}%
\bibitem [{\citenamefont {Jiang}\ \emph {et~al.}(2019)\citenamefont {Jiang},
  \citenamefont {Jiang},\ and\ \citenamefont {Chen}}]{Jiang-2019}%
  \BibitemOpen
  \bibfield  {author} {\bibinfo {author} {\bibfnamefont {S.}~\bibnamefont
  {Jiang}}, \bibinfo {author} {\bibfnamefont {Z.}~\bibnamefont {Jiang}}, \ and\
  \bibinfo {author} {\bibfnamefont {Q.}~\bibnamefont {Chen}},\ }\href@noop {}
  {\bibfield  {journal} {\bibinfo  {journal} {Sci. Rep.}\ }\textbf {\bibinfo
  {volume} {9}},\ \bibinfo {pages} {618} (\bibinfo {year} {2019})}\BibitemShut
  {NoStop}%
\bibitem [{\citenamefont {Togo}\ and\ \citenamefont {Tanaka}(2015)}]{phonopy}%
  \BibitemOpen
  \bibfield  {author} {\bibinfo {author} {\bibfnamefont {A.}~\bibnamefont
  {Togo}}\ and\ \bibinfo {author} {\bibfnamefont {I.}~\bibnamefont {Tanaka}},\
  }\href@noop {} {\bibfield  {journal} {\bibinfo  {journal} {Scr. Mater.}\
  }\textbf {\bibinfo {volume} {108}},\ \bibinfo {pages} {1} (\bibinfo {year}
  {2015})}\BibitemShut {NoStop}%
\bibitem [{\citenamefont {Bl\"{o}chl}(1994)}]{PAW-Blochl-1994}%
  \BibitemOpen
  \bibfield  {author} {\bibinfo {author} {\bibfnamefont {P.~E.}\ \bibnamefont
  {Bl\"{o}chl}},\ }\href@noop {} {\bibfield  {journal} {\bibinfo  {journal}
  {Phys. Rev. B}\ }\textbf {\bibinfo {volume} {50}},\ \bibinfo {pages} {17953}
  (\bibinfo {year} {1994})}\BibitemShut {NoStop}%
\bibitem [{\citenamefont {Kresse}(1995)}]{VASP-Kresse-1995}%
  \BibitemOpen
  \bibfield  {author} {\bibinfo {author} {\bibfnamefont {G.}~\bibnamefont
  {Kresse}},\ }\href@noop {} {\bibfield  {journal} {\bibinfo  {journal} {J.
  Non-Cryst. Solids}\ }\textbf {\bibinfo {volume} {193}},\ \bibinfo {pages}
  {222} (\bibinfo {year} {1995})}\BibitemShut {NoStop}%
\bibitem [{\citenamefont {Kresse}\ and\ \citenamefont
  {Furthm\"{u}ller}(1996)}]{VASP-Kresse-1996}%
  \BibitemOpen
  \bibfield  {author} {\bibinfo {author} {\bibfnamefont {G.}~\bibnamefont
  {Kresse}}\ and\ \bibinfo {author} {\bibfnamefont {J.}~\bibnamefont
  {Furthm\"{u}ller}},\ }\href@noop {} {\bibfield  {journal} {\bibinfo
  {journal} {Comput. Mater. Sci.}\ }\textbf {\bibinfo {volume} {6}},\ \bibinfo
  {pages} {15} (\bibinfo {year} {1996})}\BibitemShut {NoStop}%
\bibitem [{\citenamefont {Kresse}\ and\ \citenamefont
  {Joubert}(1999)}]{VASP-Kresse-1999}%
  \BibitemOpen
  \bibfield  {author} {\bibinfo {author} {\bibfnamefont {G.}~\bibnamefont
  {Kresse}}\ and\ \bibinfo {author} {\bibfnamefont {D.}~\bibnamefont
  {Joubert}},\ }\href@noop {} {\bibfield  {journal} {\bibinfo  {journal} {Phys.
  Rev. B}\ }\textbf {\bibinfo {volume} {59}},\ \bibinfo {pages} {1758}
  (\bibinfo {year} {1999})}\BibitemShut {NoStop}%
\bibitem [{\citenamefont {Perdew}\ \emph {et~al.}(2008)\citenamefont {Perdew},
  \citenamefont {Ruzsinszky}, \citenamefont {Csonka}, \citenamefont {Vydrov},
  \citenamefont {Scuseria}, \citenamefont {Constantin}, \citenamefont {Zhou},\
  and\ \citenamefont {Burke}}]{PBEsol}%
  \BibitemOpen
  \bibfield  {author} {\bibinfo {author} {\bibfnamefont {J.~P.}\ \bibnamefont
  {Perdew}}, \bibinfo {author} {\bibfnamefont {A.}~\bibnamefont {Ruzsinszky}},
  \bibinfo {author} {\bibfnamefont {G.~I.}\ \bibnamefont {Csonka}}, \bibinfo
  {author} {\bibfnamefont {O.~A.}\ \bibnamefont {Vydrov}}, \bibinfo {author}
  {\bibfnamefont {G.~E.}\ \bibnamefont {Scuseria}}, \bibinfo {author}
  {\bibfnamefont {L.~A.}\ \bibnamefont {Constantin}}, \bibinfo {author}
  {\bibfnamefont {X.}~\bibnamefont {Zhou}}, \ and\ \bibinfo {author}
  {\bibfnamefont {K.}~\bibnamefont {Burke}},\ }\href@noop {} {\bibfield
  {journal} {\bibinfo  {journal} {Phys. Rev. Lett.}\ }\textbf {\bibinfo
  {volume} {100}},\ \bibinfo {pages} {136406} (\bibinfo {year}
  {2008})}\BibitemShut {NoStop}%
\bibitem [{\citenamefont {Methfessel}\ and\ \citenamefont
  {Paxton}(1989)}]{Methfessel-1989}%
  \BibitemOpen
  \bibfield  {author} {\bibinfo {author} {\bibfnamefont {M.}~\bibnamefont
  {Methfessel}}\ and\ \bibinfo {author} {\bibfnamefont {A.~T.}\ \bibnamefont
  {Paxton}},\ }\href@noop {} {\bibfield  {journal} {\bibinfo  {journal} {Phys.
  Rev. B}\ }\textbf {\bibinfo {volume} {40}},\ \bibinfo {pages} {3616}
  (\bibinfo {year} {1989})}\BibitemShut {NoStop}%
\bibitem [{\citenamefont {Pizzi}\ \emph {et~al.}(2016)\citenamefont {Pizzi},
  \citenamefont {Cepellotti}, \citenamefont {Sabatini}, \citenamefont
  {Marzari},\ and\ \citenamefont {Kozinsky}}]{AiiDA}%
  \BibitemOpen
  \bibfield  {author} {\bibinfo {author} {\bibfnamefont {G.}~\bibnamefont
  {Pizzi}}, \bibinfo {author} {\bibfnamefont {A.}~\bibnamefont {Cepellotti}},
  \bibinfo {author} {\bibfnamefont {R.}~\bibnamefont {Sabatini}}, \bibinfo
  {author} {\bibfnamefont {N.}~\bibnamefont {Marzari}}, \ and\ \bibinfo
  {author} {\bibfnamefont {B.}~\bibnamefont {Kozinsky}},\ }\href@noop {}
  {\bibfield  {journal} {\bibinfo  {journal} {Comput. Mater. Sci.}\ }\textbf
  {\bibinfo {volume} {111}},\ \bibinfo {pages} {218 } (\bibinfo {year}
  {2016})}\BibitemShut {NoStop}%
\bibitem [{Aii(265a)}]{AiiDA-VASP}%
  \BibitemOpen
  \href@noop {} {}\bibinfo {howpublished}
  {\url{https://github.com/aiida-vasp/aiida-vasp}} (\bibinfo {year} {commit
  be4b265a})\BibitemShut {NoStop}%
\bibitem [{Aii(ad35)}]{AiiDA-phonopy}%
  \BibitemOpen
  \href@noop {} {}\bibinfo {howpublished}
  {\url{https://github.com/aiida-phonopy/aiida-phonopy}} (\bibinfo {year}
  {commit 26c3ad35})\BibitemShut {NoStop}%
\bibitem [{\citenamefont {Wood}(1962)}]{Wood-1962}%
  \BibitemOpen
  \bibfield  {author} {\bibinfo {author} {\bibfnamefont {R.~M.}\ \bibnamefont
  {Wood}},\ }\href@noop {} {\bibfield  {journal} {\bibinfo  {journal} {Proc.
  Phys. Soc.}\ }\textbf {\bibinfo {volume} {80}},\ \bibinfo {pages} {783}
  (\bibinfo {year} {1962})}\BibitemShut {NoStop}%
\bibitem [{\citenamefont {Stassis}\ \emph {et~al.}(1979)\citenamefont
  {Stassis}, \citenamefont {Arch}, \citenamefont {Harmon},\ and\ \citenamefont
  {Wakabayashi}}]{Stassis-Ti-phonon-1979}%
  \BibitemOpen
  \bibfield  {author} {\bibinfo {author} {\bibfnamefont {C.}~\bibnamefont
  {Stassis}}, \bibinfo {author} {\bibfnamefont {D.}~\bibnamefont {Arch}},
  \bibinfo {author} {\bibfnamefont {B.~N.}\ \bibnamefont {Harmon}}, \ and\
  \bibinfo {author} {\bibfnamefont {N.}~\bibnamefont {Wakabayashi}},\
  }\href@noop {} {\bibfield  {journal} {\bibinfo  {journal} {Phys. Rev. B}\
  }\textbf {\bibinfo {volume} {19}},\ \bibinfo {pages} {181} (\bibinfo {year}
  {1979})}\BibitemShut {NoStop}%
\bibitem [{\citenamefont {Ko}\ \emph {et~al.}(2015)\citenamefont {Ko},
  \citenamefont {Grabowski},\ and\ \citenamefont
  {Neugebauer}}]{Won-Seok-Ti-phonon-2015}%
  \BibitemOpen
  \bibfield  {author} {\bibinfo {author} {\bibfnamefont {W.-S.}\ \bibnamefont
  {Ko}}, \bibinfo {author} {\bibfnamefont {B.}~\bibnamefont {Grabowski}}, \
  and\ \bibinfo {author} {\bibfnamefont {J.}~\bibnamefont {Neugebauer}},\
  }\href@noop {} {\bibfield  {journal} {\bibinfo  {journal} {Phys. Rev. B}\
  }\textbf {\bibinfo {volume} {92}},\ \bibinfo {pages} {134107} (\bibinfo
  {year} {2015})}\BibitemShut {NoStop}%
\bibitem [{\citenamefont {Aroyo}\ \emph {et~al.}(2014)\citenamefont {Aroyo},
  \citenamefont {Orobengoa}, \citenamefont {de~la Flor}, \citenamefont {Tasci},
  \citenamefont {Perez-Mato},\ and\ \citenamefont {Wondratschek}}]{Aroyo-2014}%
  \BibitemOpen
  \bibfield  {author} {\bibinfo {author} {\bibfnamefont {M.~I.}\ \bibnamefont
  {Aroyo}}, \bibinfo {author} {\bibfnamefont {D.}~\bibnamefont {Orobengoa}},
  \bibinfo {author} {\bibfnamefont {G.}~\bibnamefont {de~la Flor}}, \bibinfo
  {author} {\bibfnamefont {E.~S.}\ \bibnamefont {Tasci}}, \bibinfo {author}
  {\bibfnamefont {J.~M.}\ \bibnamefont {Perez-Mato}}, \ and\ \bibinfo {author}
  {\bibfnamefont {H.}~\bibnamefont {Wondratschek}},\ }\href@noop {} {\bibfield
  {journal} {\bibinfo  {journal} {Acta Crystallogr. Sect. A}\ }\textbf
  {\bibinfo {volume} {70}},\ \bibinfo {pages} {126} (\bibinfo {year}
  {2014})}\BibitemShut {NoStop}%
\bibitem [{\citenamefont {Dove}(1993)}]{IntroductionToLatticeDynamics}%
  \BibitemOpen
  \bibfield  {author} {\bibinfo {author} {\bibfnamefont {M.~T.}\ \bibnamefont
  {Dove}},\ }\href@noop {} {\emph {\bibinfo {title} {Introduction to Lattice
  Dynamics}}}\ (\bibinfo  {publisher} {Cambridge University Press},\ \bibinfo
  {year} {1993})\BibitemShut {NoStop}%
\end{thebibliography}%
\end{document}